\documentclass[aps,prl,onecolumn]{revtex4}
\usepackage{inputenc}
\usepackage[dvips]{graphicx}
\usepackage{epsfig}
\usepackage{dcolumn}
\usepackage{bm}
\usepackage{amssymb}

\begin{document}

%\title{Composite Random Walks can optimally solve the exploration-exploitation search tradeoff by adapting to the domain scales}
\title{First-passage times in multi-scale random walks: the impact of movement scales on search efficiency} 

\author{Daniel Campos$^{1}$, Frederic Bartumeus$^{2,3}$, E.P. Raposo$^{4}$ and Vicen\c{c} M\'{e}ndez$^{1}$}
\affiliation{$^{1}$Grup de F\'{\i}sica Estad\'{\i}stica, Departament de F\'{\i}sica. Universitat Aut\`{o}noma de Barcelona, 08193 Bellaterra
(Barcelona) Spain.\\
$^{2}$ICREA Movement Ecology Laboratory (CEAB-CSIC) Cala
Sant Francesc 14, 17300 Blanes, Girona, Spain. \\
$^{3}$ CREAF, 08193 Bellaterra (Barcelona) Spain\\
$^{4}$ Laborat\'{o}rio de F\'{i}sica Te\'{o}rica e Computacional,
Departamento de F\'{i}sica, Universidade Federal de Pernambuco, Recife-PE, Brazil}

\begin{abstract}

An efficient searcher needs to balance properly the tradeoff between the exploration of new spatial
areas and the exploitation of nearby resources, an idea which is at the core of \textit{scale-free} L\'evy search
strategies. Here we study multi-scale random walks as an approximation to the scale-
free case and derive the exact expressions for their mean-first passage times in a one-dimensional
finite domain. This allows us to provide a complete analytical description of the
dynamics driving the asymmetric regime, in which both nearby and faraway targets are available to
the searcher. For this regime, we prove that the combination of only two movement scales can be
enough to outperform both balistic and L\'evy strategies. This two-scale strategy involves an optimal
discrimination between the nearby and faraway targets, which is only possible by adjusting
the range of values of the two movement scales to the typical distances between encounters. So,
this optimization necessarily requires some prior information (albeit crude) about targets distances or distributions.
Furthermore, we found that the incorporation of additional (three, four, ...) movement scales and its adjustment to target distances does not improve further the
search efficiency. This allows us to claim that optimal random search strategies in the asymmetric regime
actually arise through the informed combination of only two walk scales (related to the exploitative
and the explorative scale, respectively), expanding on the well-known result that optimal strategies in strictly
uninformed scenarios are achieved through L\'evy paths (or, equivalently, through a hierarchical combination of multiple scales).

\end{abstract}

\maketitle

%Random search theory can be seen as a branch of stochastic processes
%which tries to answer two questions: (i) Given an individual moving
%according to a random-walk scheme, what time will it take until it
%successfully \textit{detects} a certain number of targets whose
%locations are unknown (or only partially known)? (ii) And how could
%this random searcher regulate its random path in order to optimize
%such search efficiency? Search path optimizaiton has been typically
%addressed in the physical literature through the computation, and
%subsequent optimization, of the Mean-First Passage Time (MFPT) of
%random walkers \cite{viswanathan11,benichou11,monasterio11,mendez14},
%a magnitude that is not only of interest in search problems but also
%in many areas of physics and science in general
%\cite{redner01,metzler14}.

%The existence  of the  unexpected $\mu \approx  2$ optimality  for the
%\textit{asymmetric} regime has had  a notable impact on the ecological
%community, and has lead to  a controversy about the possible emergence
%of  L\'{e}vy  (scale-free)  signatures   in  the  analysis  of  animal
%trajectories  and  its  subsequent   interpretation  in  terms  of  an
%evolutionary  adaptation  driven  by  search  efficiency  maximization
%\cite{viswanathan11,bartumeus09,james11}. 

%Where we introduce random search theory and current state-of-the art
Search  theory aims  at identifying  optimal strategies  that help  to
promote  encounters between  a searcher  and its  target.  Statistical
physics approaches often identify  searchers as  random walkers, capitalizing  on the
idea of search as movement under uncertainty.  The assumption that the
searcher lacks  any information  about target  locations leads  to the
fundamental question of how individual paths should be orchestrated to
enhance              random               encounter              rates
\cite{viswanathan11,benichou11,monasterio11,mendez14}.  Based on these
assumptions, a  proper measure  of search efficiency  is given  by the
Mean-First Passage Time (MFPT) of the random walker through the target
location, a quantity which is also the focus of interest in many other
areas of physics and science \cite{redner01,metzler14}.  Random search
theory was spurred  in the nineties as  a result of a  series of works
suggesting that  L\'{e}vy patterns (particularly, L\'{e}vy  walks) can
be  optimal strategies  for uninformed  space exploration  (see, e.g.,
\cite{shlesinger86,cole95,schuster96,viswanathan96,berkolaiko96,viswanathan99}),
and it has continually been developed since then.

L\'{e}vy  walks   can  be   defined  as   random  paths   composed  of
statistically identical flights  whose length probability distribution
decays  asimptotically   as  $\sim   \vert  x  \vert   ^{-\mu}$  (with
$1<\mu<3$), where  $\mu$ is  frequently known as  the \textit{L\'{e}vy
  exponent}. So,  lower values of the  L\'{e}vy exponent comparatively
imply a  higher frequency  of long flights.   As a  remarkable result,
Viswanathan   and  colleagues   \cite{viswanathan99}  identified   two
different regimes  associated to  random search dynamics  depending on
whether  targets   are  completely  or  uncompletely   depleted  after
encounter,      the       so-called      \textit{destructive}      and
\textit{non-destructive}   dynamics,   respectively.   More   recently
\cite{mendez14,bartumeus14}  it has  been  emphasized  that these  two
dynamical regimes are much general  and should be renamed and directly
associated to searcher-to-target distances.  In the \textit{symmetric}
regime  targets are  expected  to occur  at  an average,  characteristic
distance from the searcher.  For example, a
\textit{destructive} search  dynamics tend  to deploy  targets locally
and promote targets  being faraway (on average) from  the searcher, at
least in low-density and  homogeneous landscapes. A \textit{symmetric}
regime may also  correpond to high target density  scenarios where (on
average)  most  targets  can  be   assumed  to  be  closeby.   In  the
\textit{asymmetric},  instead, a  wide  variety of  searcher-to-target
distances exist  (i.e.  heterogeneous  landscapes), and both  near and
faraway targets may coexist in different proportions.
 
 A more  convenient understanding and interpretation  of these regimes
 can    be   attained    if    linked   to    the   general    binomia
 exploitation-exploration \cite{mendez14}.  According to it, three
 different scenarios should be  distinguished: (i) those situations in
 which  exploration  is  clearly  preferred over  exploitation  (so  a
 ballistic strategy, defined as a straight trajectory without changes of direction, is then trivially  expected to be optimal). This is
 the case  where revisiting  areas is  worthless and  the optimization
 requires  performing  displacements  as   long  as  possible  without
 changing  direction; a  ballistic strategy  is thus  preferable. This
 happens  if  targets  are  uniformly distributed  and  can  be  fully
 depleted  or  if  all  targets  are faraway  (on  average)  from  the
 searcher.   (ii) those  situations where  exploitation prevails  over
 exploration (so  local, spatially  bounded or diffusive  search is
 optimal).   This is  the case  of a  searcher being  nearby a  set of
 targets (patch)  that is never  depleted. The random  searcher always
 has the  possibility to come  back to the  patch and the  strategy of
 sticking around  it is much  preferable because no other  targets are
 available in the landscape.  And  (iii) those situations where a true
 exploitation-exploration   tradeoff   emerges  because   the   search
 necessarily requires  the ability  to reach  both nearby  and distant
 targets.  For example  if  target distribution  is  patchy and  while
 walking one can have nearby and faraway patches.

%Where we explain the problem of MFPTs aproximations in complex, superdiffusive stochastic processes
While  the   dynamics  in   the  \textit{symmetric}  regime   is  thus
straightforward to understand in terms of maximization (scenario i) or
minimization (scenario ii)  of the area explored,  the details driving
optimization  in the  \textit{asymmetric}  regime  (scenario iii),  in
particular  how  movement  scales determine  search  efficiency,  have
remained partially  obscure to date  \cite{bartumeus14,mendez14}. This
is so  because analytical  methods for the  determination of  the MFPT
(often valid just for Markovian  processes) are difficult to extend to
L\'{e}vy or  other superdiffusive  dispersal mechanisms.  In  the last
years, much effort has been  devoted to overcome this limitation.  For
instance,   in  \cite{koren07}   the  asymptotic   behaviour  of   the
first-passage distribution of L\'{e}vy  flights in semi-infinite media
was  obtained.  Other  authors  have derived  expressions and  scaling
properties  of   MFPTs  for  moving  particles   described  either  by
Fractional  Brownian  Motion \cite{sanders12,malley11}  or  fractional
diffusion  equations  \cite{gitterman00,dybiec06,dybiec10}.   Finally,
the  alternative approach  to  approximate L\'{e}vy  paths through  an
upper bound  truncation (so  that L\'evy properties  hold just  over a
specific set of scales) has  been explored too \cite{bartumeus14}. But
despite  these  advances, analytical  arguments  able  to explain  the
different  optimization dynamics  observed in  the \textit{asymmetric}
compared to the \textit{symmetric} regime are still lacking.

 %The initial distance
 %of  the searcher  to the  nearest target  becomes then  the essential
 %agnitude     to    quantify     the    \textit{symmetric-asymmetric}
 %transition. 

%utility of multi-scale random walks, and connection to biology.
L\'{e}vy or scale-free paths  can be conveniently approximated through
a combination of multiple  scales \cite{hughes81}.  This is tantamount
to  expressing power-law  functions as  a combination  of exponentials
\cite{reynolds14,goychuk09}  or  providing  a Markovian  embedding  for  L\'{e}vy
stochastic     processes    \cite{kupferman04,siegle10,lubashevsky09}.
Composite random  walks and, more in general, Multi-Scale Random Walks (MSRW)  have  also  emerged  recently  as  an
alternative  to  the  presence  of  scale-free  signatures  in  animal
trajectories  \cite{reynolds14,benhamou14}.   It   is  not  clear  yet
whether the  emergence of  multi-scaled movement behaviour  in biology
responds to exploratory behaviour tuned  to uncertainty (L\'evy as the
limiting  case)  \cite{bartumeus14,mendez14},   or  else  to  informed
behavioural  processes  linked  to landscape  through  sensors  and/or
memory.   It is  thus important  to understand  how this  multi-scaled
behaviour should  be coupled with other  relevant landscape magnitudes
like target distributions and searcher-to-target average distances.

Inspired  by these  ideas,  in the  present work  we  derive an  exact
analytical method  for the  determination of  the MFPT  of MSRWs  as an
approximation  to  the  scale-free  case.
While  the method  proposed becomes  increasingly complicated  as more
scales  are  considered,  we  show   that  2-scale  random  walks  can
effectively resolve  the explotation-exploration tradeoff  emergent in
the \textit{asymmetric} regime by  adjusting movement scales to target
distances.  Furthermore,  the comparison  between the 2-scale  and the
3-scale random walk suggests that incorporating a third scale does not
produce any  advantage.
%Això no és veritat
%, unless the  3-scales are indeed  reproducing a
%Weierstrassian   walk,   thus,   approximating  a   L\'evy   strategy.
Therefore, we conclude  that an optimal random search  strategy in the
\textit{asymmetric} regime consists on combining two informed movement
scales that  should approximately correspond to  nearby/faraway target
distances.  Hence, an informed adjustement of movement scales improves
search efficiency compared to  any non-informed strategy (where scales
are  imposed at  random).   In the  case  of non-informed  strategies,
however,  MSRWs aproximating  the  L\'evy  strategy
are   the  best  solution   to  solve exploitation-exploration tradeoffs.

\section{Derivation of the MFPT}

We consider for simplicity an isotropic random
walk  embedded  in  the  one-dimensional finite  domain  $(0,L)$  with
initial  position  $x_{0}$  and  absorbing boundaries  (so  implicitly
assuming that  surrounding targets  are located at  $x=0$ and  $x=L$). We will choose $x_{0}<L/2$ arbitrarily, so $x_{0}$ can be interpreted as the initial distance of the searcher to the nearest target. The
searcher  moves  continuously  with  constant  speed  $v$  and  performs
consecutive  flights  whose duration is distributed   according  to  a  multiexponential
distribution function $\varphi(t)$ in the form
\begin{equation}
\varphi(t)= \sum_{i=1}^{n} w_{i} \varphi_{i} (t), \quad  \varphi_{i}(t)= \tau_{i}^{-1} e^{-t/\tau_{i}},
\label{varphi}
\end{equation}
so yielding  a $n$-scale MSRW  characterized by the  persistence times
$\tau_{i}$  and their  corresponding weights  $w_{i}$, that  satisfy the
normalization condition $\sum_{i=1}^{n} w_{i}=1$.

We now define $\rho_{i}(x,t;\boldsymbol{\rho}_{0})$ as the probability
that  the  walker  starts  at  time  $t$  from  $x$  a  single  flight
characterized by the  distribution $\varphi_{i}(t)$ (in the following,
using  a   particular  distribution  $\varphi_{i}(t)$   is  termed  as
\textit{being  in  state}  $i$).  The  vector  $\boldsymbol{\rho}_{0}=
\delta(x-x_{0})  (\rho_{10},\rho_{20},\ldots,\rho_{n0})$  accounts for
the  set of  initial conditions  in all  states, with  $\rho_{i0}$ the
probability of being  in state $i$ at $t=0$.  Using this notation, the
multi-scale  (non-Markovian)  walk  gets  reduced  to  a  set  of  $n$
Markovian states which satisfy (according to standard prescriptions of
the  Continuous-Time  Random   Walk  \cite{mendez14})  the  mesoscopic
balance equations
\begin{equation}
\rho_{i}(x,t;\boldsymbol{\rho}_{0})=w_{i} \sum_{k=1}^{n} \int_{0}^{t} \left( \frac{ \rho_{k+} + \rho_{k-} }{2} \right) \varphi_{k}(t') dt'  +\rho_{i0} \delta(x-x_{0})\delta(t)
\label{rho}
\end{equation} 
(for   $i=1,2,\ldots,n$),  where  we   have  introduced   the  compact notation
$\rho_{i\pm}  \equiv \rho_{i}(x  \pm vt',t-t';\boldsymbol{\rho}_{0})$.
The corresponding probability that  the walker, passing through $x$ at
time $t$, is performing at that instant a flight in state $i$ is given
by
\begin{equation}
P_{i}(x,t;\boldsymbol{\rho}_{0})= \int_{0}^{t} \left( \frac{ \rho_{i+} + \rho_{i-} }{2} \right) \tau_{i} \varphi_{i}(t') dt' .
\label{P}
\end{equation}
Here we have used the relation $\int_{t}^{\infty} \varphi_{i}(t') dt'=
\tau_{i} \varphi_{i}(t)$,  valid for exponential  distributions, which
gives the probability  that a single flight in state  $i$ will last at
least a time $t$.

Due to  the Markovian  embedding used, the  general propagator  of the
random walk in  an infinite media can be written  in the Laplace space
(with   $s$    the   Laplace   argument)    as   $\sum_{i}^{n}   P_{i}
(x,s;\boldsymbol{\rho}_{0})$ with the  probability density for state $i$,  $P_{i}(x,t;\boldsymbol{\rho}_{0})$, given  by a sum  of $n$
exponentials
\begin{equation}
P_{i} (x,s;\boldsymbol{\rho}_{0}) = \sum_{j=1}^{n} \alpha_{ij}(s) e^{-\beta_{j}(s) \vert x-x_{0} \vert /v},
\label{infinite}
\end{equation}
where $\alpha_{ij}$ and $\beta_{j}$ are positive constants to be determined from the solution of the system (\ref{varphi}-\ref{P}). Hence, the solution in the interval of interest $(0,L)$ with periodic boundary conditions reads
\begin{eqnarray}
\nonumber Q_{i}(x,s;\boldsymbol{\rho}_{0}) &\equiv & \sum_{m=-\infty}^{\infty} P_{i}(x+mL,s;\boldsymbol{\rho}_{0}) = \\ &=& \sum_{j=1}^{n} \alpha_{ij}(s) \frac{ e^{-\beta_{j}(s)(L-\vert x-x_{0} \vert)/v} e^{-\beta_{j}(s) \vert x-x_{0} \vert /v}} {1-e^{-\beta_{j}(s) L/v}}.
\label{periodic}
\end{eqnarray}

Finally, the exact  MFPT can be computed from  Eq. (\ref{periodic}) by
extending  known methods  for Markovian  processes; in  particular, we
employ      here     the      renewal     method      for     velocity
models \cite{campos12,campos14}.    According   to   this,  we   define
$f_{i}(t;\boldsymbol{\rho}_{0})$ as the first-passage time probability
rate for a  walker through any of the boundaries  while being in state
$i$. The renewal property of  Markovian processes allows then to write
the recurrence relations
\begin{equation}
q_{i}(t;\boldsymbol{\rho}_{0})=f_{i}(t;\boldsymbol{\rho}_{0}) + \sum_{k=1}^{n} \int_{0}^{t} f_{k}(t-t';\boldsymbol{\rho}_{0}) q_{i} (t';\boldsymbol{\rho}_{k}) dt',
\label{fpt}
\end{equation}
where $q_{i}(t;\boldsymbol{\rho}_{0})$  is defined as  the probability
rate with which  the walker hits (not necessarily  for the first time)
the  boundary  at  time  $t$  while  being  in  state  $i$.  The  term
$q_{i}(t;\boldsymbol{\rho}_{k})$ has the same meaning but for a walker
starting  its path  at state  $k$ from  the boundary  (so with $x_{0}=0$). According       to       (\ref{fpt})       the      hitting       rate
$q_{i}(t;\boldsymbol{\rho}_{0})$ gets  divided into those trajectories
for     which      this     is     the      first     hitting     rate
($f_{i}(t;\boldsymbol{\rho}_{0})$)  plus those  trajectories  that hit
the boundary for the first time  at a previous time $t-t'$ in any of the possible
$n$  states  (second  term   on  the  lhs of (\ref{fpt})).  The  total  first-passage
distribution   of   the  MSRW   will  read   then
$f(t;\boldsymbol{\rho}_{0})=\sum_{i=1}^{n}
f_{i}(t;\boldsymbol{\rho}_{0})$  (where   the  $f_{i}$'s  are   to  be
determined from the system  of equations (\ref{fpt})), and the general
expression for the MFPT will be by definition \cite{mendez14}
\begin{equation}
\langle T \rangle = \lim_{s \rightarrow 0} \sum_{i=1}^{n} \frac{ d f_{i}(s;\boldsymbol{\rho}_{0}) } {ds}.
\label{mfpt}
\end{equation}

Then, to  find a  closed expression for  $\langle T \rangle$  one just
needs  to express  the hitting  rates $q_{i}(t;\boldsymbol{\rho}_{0})$
and $q_{i}(t;\boldsymbol{\rho}_{k})$ in terms  of the solutions of the
random-walk (\ref{varphi}-\ref{periodic}).  This is given,  in analogy
to previous works \cite{verechtchaguina06,campos14}, by
\begin{eqnarray}
\nonumber q_{i}(t;\boldsymbol{\rho}_{0}) &=& \left\{ 
\begin{array}{cc}
v Q_{j}(0,t;\boldsymbol{\rho}_{0}),   & 0<x_{0}<L \\ 
v Q_{j}(0,t;\boldsymbol{\rho}_{0}) - \delta (t) /2, & \quad x_{0}=0, x_{0}=L ;
\end{array}
\right.  \\
q_{i}(t;\boldsymbol{\rho}_{k}) &=& v Q_{j}(0,t;\boldsymbol{\rho}_{k}) - \delta (t) /2 .
\label{rate}
\end{eqnarray}

Here clearly a different behaviour for the case when the walker starts
from the boundaries is introduced by convenience to make explicit that
the walker cannot get trapped by the  boundary immediately at $t=0$,
but hittings  are only possible for  $t>0$. In the  following we study
how the different  scales considered in the MSRW contribute to
the  search efficiency as a function of the two prominent spatial scales present in the problem (i.e. $x_{0}$ and $L$). 

\subsection{1-scale case ($n=1$)} 

The MSRW scheme described above reduces
trivially in this case to a classical Correlated Random Walk (see,
e.g.,  \cite{goldstein51,mendez14,campos15}) for which the free propagator (Equation (\ref{infinite})) reads
\begin{equation}
P_{1} (x,s;\boldsymbol{\rho}_{0}) =\frac{1}{2v} \sqrt{ \frac{s + \tau_{1}^{-1}}{s} } \mathrm{exp} \left[ \sqrt{ s \left( s+ \tau_{1}^{-1} \right) } \vert x-x_{0} \vert / v \right]
\end{equation}

Using  the derivations  in Equations (\ref{periodic}-\ref{rate}), the
MFPT  in (\ref{mfpt}) yields  the exact  expression obtained  by Weiss
\cite{weiss84} thirty years ago
\begin{equation}
\langle T \rangle = \frac{L}{2v} + \frac{ x_{0} (L-x_{0})} {v^{2} \tau_{1}} .
\label{weiss}
\end{equation}

So,  assuming  that  $x_{0}$,  $L$   are  fixed  by  the  external  or
environmental conditions,  we observe that the  search optimization of
the 1-scale random walk turns out to be trivial: faster searches (i.e.
larger  values of  $v$) and  straighter trajectories  (i.e.  $\tau_{1}
\rightarrow \infty$)  will monotonically  reduce the search  time.  In
particular, note  that for $\tau_{1} \rightarrow  \infty$ one recovers
the result $\langle T \rangle  =L/2v$, which coincides with the result
for a  ballistic strategy.  It  is clear  then that in  1-scale random
walks the  exploration-exploitation tradeoff ($L$ vs  $x_0$) is
always trivially  optimized through a  ballistic strategy (in agreement
with  the results  in \cite{bartumeus14}).   As  we shall  see in  the
following,  at least  2 scales  are necessary  in the  random walk  to
observe such effects.

%DANI: The fact that 1-scale RW can not optimize correctly the asymmetric regimes does not mean the 
%asymmetric regime does not exist right? It is simply that it needs to pick one scale and that is it....
%So the sentence above is not very clear to me
%FEDE: Aquest comentari anava en la línia de definir règim asimètric com un regim matemàtic.
%Si defineixes asymmetric com una situació o regim biologic, aleshores efectivament no té massa sentit.

\subsection{2-scales case ($n=2$)} 

The exact analytical solution for this case can still be found  easily, albeit the general expression for the
MFPT obtained is cumbersome; details are provided in the Supplementary Information (SI) file.  A first survey on this solution (which was implemented in MAPLE) allows us to observe that for \textit{large} values of $x_{0}$ the balistic-like strategy (i.e. $\tau_{1} \rightarrow \infty$, $\tau_{2} \rightarrow \infty$) is again the one which minimizes $\langle T \rangle$. However, for \textit{small} $x_{0}$ values we find now the emergence of an asymmetric regime in which the optimal is attained for one of the two scales (either $\tau_{1}$ or $\tau_{2}$) being much larger than the time $L/v$ required to cover the domain, with the other scale exhibiting a smaller value. The  threshold at which this transition occurs (so, the value of $x_{0}$ for which the optimum $\langle T \rangle$ becomes smaller than $L/2v$) turns out to be  $x_{0} \approx 0.105L$, a value which is confirmed by random-walk simulations too.

At the sight of these results, we will focus now our interest in providing some limit expressions which can help us to understand how this transition occurs and how the system behaves in the \textit{asymmetric} regime. First we note that, in solving the exploration-exploitation tradeoff, the exploration part will be always optimized through flights much longer than the typical time to cover the whole domain, which explains why one of the two scales (say, $\tau_{1}$) should be expected to be as large as possible, in particular $\tau_{1} \gg L/v$. Regarding the second scale, the exploitation side of the tradeoff (corresponding to exploring the surrounding area searching for nearby targets) should intuitively benefit from choosing a scale of the order of $\tau_{2} \sim x_{0}/v$, the time required to reach the nearest target. Scales much larger than this would promote exploration instead of exploitation, while scales much smaller would lead to an unnnecessary overlap of the searcher's trajectory around its initial position \cite{bartumeus14}. Since the \textit{asymmetric} regime must emerge necessarily from the asymmetric condition $x_{0} \ll L$ we can thus consider  that this  second  scale should satisfy $\tau_{2} \ll L/v$. 

Taking  the  two limits ($\tau_{1} v/L \rightarrow \infty$ and $\tau_{2} v/L \rightarrow 0$) into
account, our general solution for the MFPT reduces to
\begin{equation}
\langle T \rangle = \frac{L}{2v} + \frac{\tau_{2}(1-w_{1})}{w_{1}} \left( 1- \frac{ 1+\frac{L w_{1}}{2v \tau_{2}}  } {1+\sqrt{w_{1}}} \mathrm{exp}  \left[  \frac{-\sqrt{w_{1}}x} {v  \tau_{2}}  \right] \right) .
\label{mfpt2}
\end{equation}

Visual inspection of this expression already shows that values of the  MFPT below the ballistic threshold $\langle T  \rangle =  L/2v$  can  be  now  obtained  for appropriate  combinations  of  $x_{0}$,  $\tau_{2}$ and  $w_{1}$.   In particular, in the limit when $x_{0} \rightarrow 0$ the previous expression gets minimized for the  value 
\begin{equation}
\tau_{2}^{*}=\sqrt{Lx_{0}w_{1}/2v^{2}}
\label{optimal0}
\end{equation}
where we use the asterisk  to denote values that are \textit{optimal}.
After minimizing $\langle  T \rangle$ also with respect  to $w_{1}$ we
find that the global optimum of the MFPT corresponds to
\begin{equation}
\tau_{2}^{*}=\frac{1}{v} \sqrt{\frac{x_{0}^{2}(L+\sqrt{8Lx_{0}})}{L-8x_{0}}}, \quad
w_{1}^{*}=\frac{2x_{0}(L+\sqrt{8Lx_{0}})}{L(L-8x_{0})} .
\label{optimal}
\end{equation}

Now, in  the  limit  $L  \rightarrow  \infty$  we  observe  that  $\tau_{2}^{*}
\rightarrow x_{0}/v$ and $w_{1}^{*} \rightarrow 2x_{0}/L$. Altogether,
these results provide  a clear and simple description of the  search dynamics in
the  \textit{asymmetric}  regime for 2-scale MSRWs which confirms our discussion above. The  optimum  strategy in the \textit{asymmetric} regime will combine  a very large scale $\tau_{1} \gg L/v$ (for exploration purposes) with a  shorter scale $\tau_{2}$ of  the order  of $x_{0}/v$ (for better exploitation of the nearest target). It  is particularly interesting  that the optimal weight  $w_{1}^{*}$ must be
rather small, so the  searcher just needs occasional ballistic flights
while  spending  the  rest  of  the  time  searching  intensively  its
surroundings. So, the optimal strategy does not consist just on an appropriate choice of the scales $\tau_{1}$ and $\tau_{2}$ but also on using them in an adequate proportion.

\begin{figure}
\includegraphics[scale=1.0]{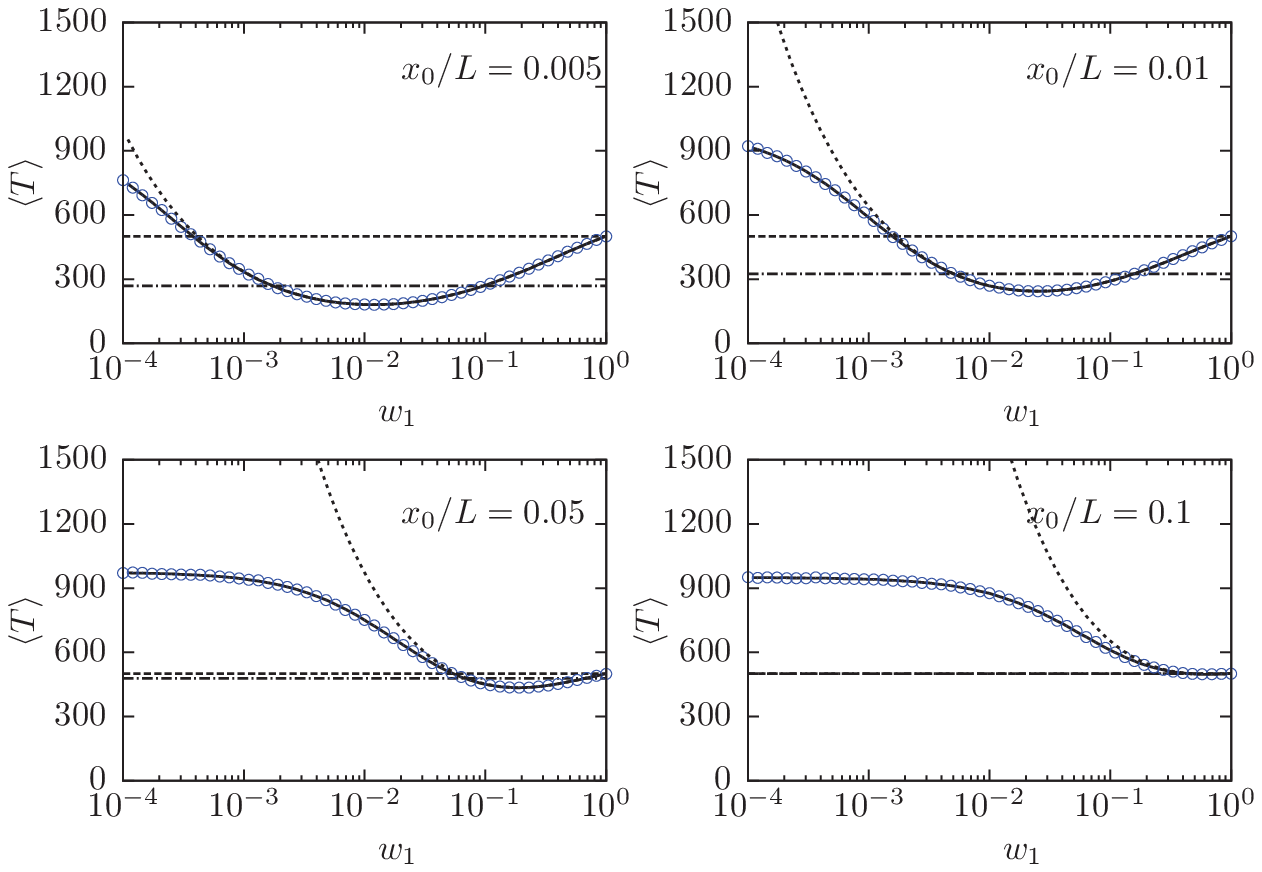}
 \caption{MFPT  for  a 2-scale  random  walker  with $L=1000$,  $v=1$,
   $\tau_{1} = 10^{3} L/v$  and  $\tau_{2}=x_{0}/v$,  and  for
   different initial conditions. The  plot shows the  exact analytical
   solution  (solid  lines),   random-walk  simulations  averaged  over
   $10^{6}$  realizations  (circles) and  the  $L \rightarrow  \infty$
   approximation  given  by Eq.  (\ref{mfpt2})  (dotted lines). The  values
   obtained  for ballistic  and $\mu=2$  L\'{e}vy strategies  are also
   given for comparison (dashed and dashed-dotted horizontal lines, respectively).}
\label{fig1}
\end{figure}

\begin{figure}
\includegraphics[scale=1.0]{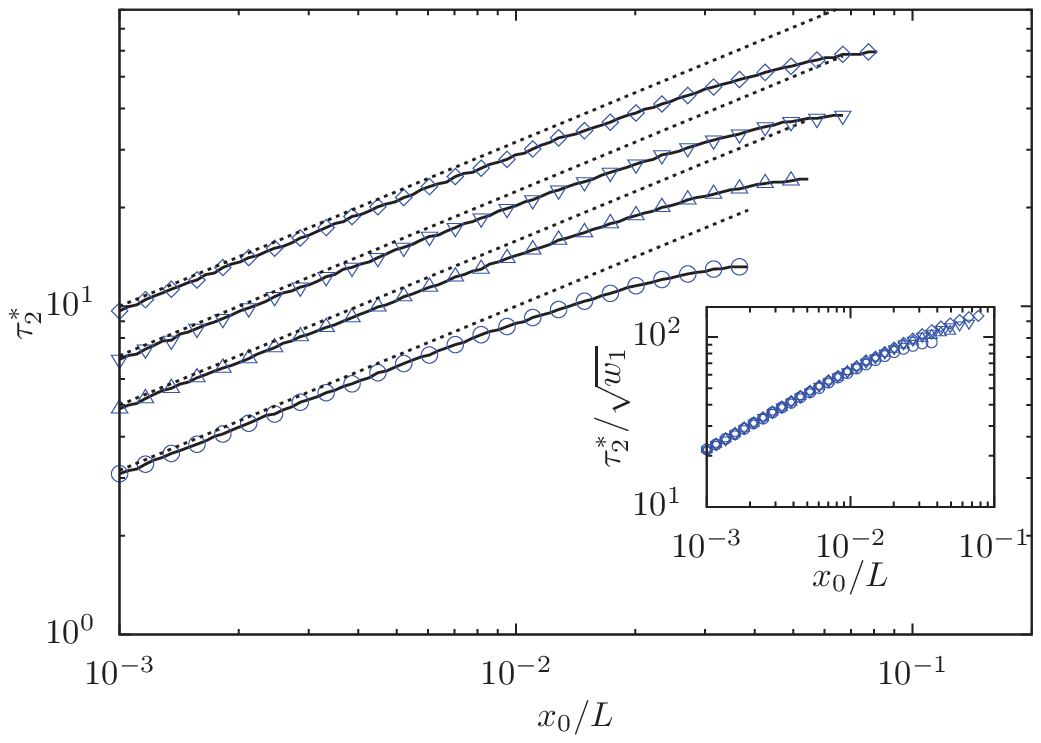}
 \caption{Optimal persistence predicted by the exact analytical solution (solid lines) and the approximated expression $\tau_{2}^{*}=\sqrt{Lx_{0}w_{1}/2v^{2}}$ (dotted lines) in comparison to random-walk simulations (symbols). Different values of the weight $w_{1}$ are shown: $w_{1}=0.02$ (circles), $0.05$ (triangles), $0.1$ (inverted triangles) and $0.2$ (diamonds). Full symbols denote the$x_{0}/L$ values above which the \textit{symmetric} regime appears and so there is no optimal persistence $\tau_{2}^{*}$. Inset: The same results are shown with $\tau_{2}^{*}/\sqrt{w_{1}}$ in the vertical axis. The collapse observed confirms the scaling $\tau_{2}^{*} \sim \sqrt{w_{1}}$ analytically derived.}
 \label{fig2}
\end{figure}

Figures  \ref{fig1}   and  \ref{fig2}  show  the   comparison  between
random-walk simulations (symbols)  and our method, both  for the exact
case     (solid     lines)      and     for     the     approximations
(\ref{mfpt2}-\ref{optimal}) (dotted lines).  Note in Figure \ref{fig1}
that the  optimum value  of the MFPT  clearly improves  (specially for
$x_{0}/L$ very small) the value obtained for a ballistic strategy or a
$\mu=2$ L\'{e}vy  walk strategy  (dashed and  dashed-dotted horizontal
lines, respectively), so revealing that an appropriate combination of
only two  move length  scales can  be actually  more efficient  than a
scale-free strategy. The range of validity of the approximated results
(\ref{mfpt2}-\ref{optimal}) is also shown in the plots, as well as the
scaling $\tau_{2}^{*} \sim  \sqrt{w_{1}}$ derived in Eq. \ref{optimal0}
(see Figure \ref{fig2}).

\begin{figure}
\includegraphics[scale=1.0]{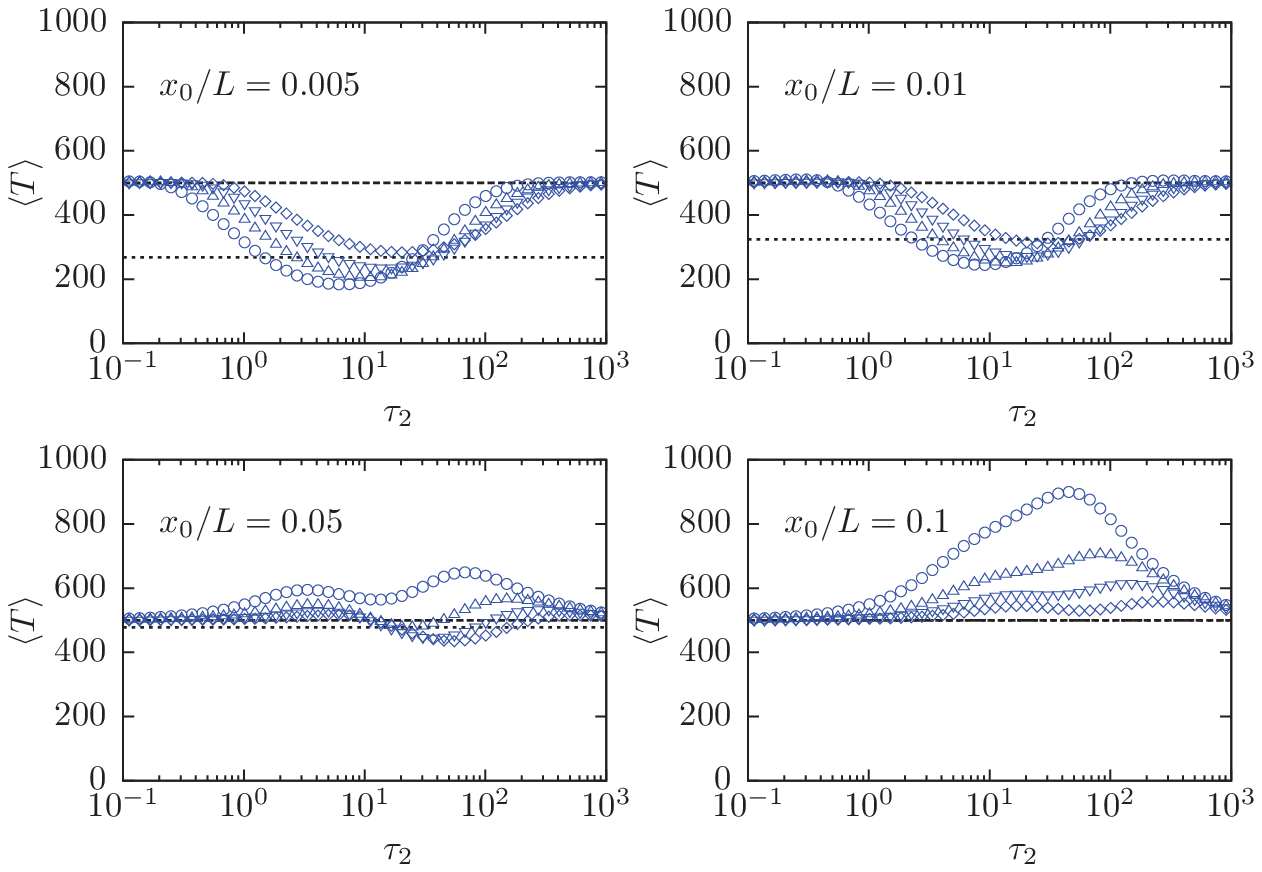}
 \caption{MFPT for a 2-scale random walker with $L=1000$, $v=1$ and $\tau_{1} = 10^{3} L/v$ with different initial conditions. The plot shows the exact analytical solution for different values $w_{1}=0.02$ (circles), $0.05$ (triangles), $0.1$ (inverted triangles) and $0.2$ (diamonds). The values obtained for ballistic and $\mu=2$ L\'{e}vy strategies are also given for comparison (dashed and dotted lines, respectively).}
\label{fig3}
\end{figure}

Despite  finding  a set  of  combinations  of $\tau_{2}$  and  $w_{1}$
outperforming  both L\'evy  and  ballistic  strategies, these  results
show that it  is necessary  for the  searcher to  have some
information about the domain scales (i.e.   $x_{0}$ and $L$) in  order to
fine-tune search and get effective strategies.  Without this knowledge
L\'evy or ballistic patterns look  as robust strategies, that could be
even  more  effective  than  searching with  badly  adjusted  movement
scales, as  suggested by  the comparison  in Figure  \ref{fig1}.  This
fact is  also confirmed  when observing the  dependence of  $\langle T
\rangle$ on $\tau_{2}$  (Figure \ref{fig3}, see also Figure  S1 in the
SI) in order to assess the range width at which $\tau_{2}$ and $w_{1}$
lead to  optimality.  In Figure  \ref{fig3} we provide the  results of
our  exact solution  for  different values  of  $x_{0}$ and  different
weights $w_{1}$ (here the approximated results and simulations are not
shown  in order  to facilitate  understanding). 
In  accordance to our analytical results above, we  observe that values of $w_{1}$  close to $2x_{0}
/L$ minimize the MFPT. So, there are certain values of $w_{1}$ for which the MFPT becomes lower than the balistic value $L/2v$ (but we we stress that the most critical parameter for getting below this threshold is clearly  $x_{0}$). Actually, for the two
upper panels (which correspond  to $x_{0}/L=0.005$ and $x_{0}/L=0.01$)
we observe that any choice of $\tau_{2}$ and $w_{1}$ would result in a
better (or as good as) performance than a balistic strategy, while the
region  where  the L\'{e}vy  strategy  is  outperformed is  relatively
small.

We stress finally that we  have carried out studies, both analytically
and numerically, for the case when the initial position $x_{0}$ is not
fixed but  is distributed  according to an  exponential or  a Gaussian
distribution, so  a range  of $x_{0}$ values  is allowed  (results not
shown here).  The  results for all these  cases coincide qualitatively
with  those  reported above,  so  whenever  values $x_{0}<0.105L$  are
predominant  the   \textit{asymmetric}  regime  is   recovered.   This
confirms that the  emergence of the \textit{asymmetric}  regime is not
an artefact  caused by  the choice  of a  fixed initial  condition, in
agreement with recent numerical studies \cite{raposo11}.

\subsection{3-scales case ($n=3$)} 

Provided that the initial time to reach any of the targets is given by
the  two   timescales  $x_{0}/v$   and  $(L-x_{0})/v$,  it   could  be
intuitively expected that  these are the only ones  necessary to reach
an optimal  strategy.  To check  this we have solved  analytically the
3-scale  case (see  SI) and,  given  that the  expression obtained  is
extremely cumbersome, we have used Markov Chain Monte Carlo algorithms
in order to determine numerically the  global minimum of the MFPT as a
function of  the parameters  $\tau_{i}$ and  $w_{i}$.  The  results so
obtained  are conclusive  and confirm  the idea  that indeed  only two
scale are needed to minimze the MFPT. We find that for large values of
$x_{0}$  the  optimal  strategy   is  again  ballistic-like  (so  only
displacements with $\tau_{i} \gg L/v$  should be performed in order to
minimize $\langle T \rangle$).  Instead,  for $x_{0}$ small enough the
optimum  arises  through the  combination  of  only two  scales  which
coincide with  those found  for the optimal  2-scale case;  this means
that two of the three scales involved (say, $\tau_{1}$ and $\tau_{2}$)
will eventually  have the  same value  after minimization.   Even more
surprising  is that  when the  initial  condition is  governed by  two
different scales (by  combining two different values  of $x_{0}$, each
with a given  probability) the optimum still corresponds  to a 2-scale
random-walk;  in this  case  the optimum  value  $\tau_{2}^{*}$ is  in
between the  optimum values that  one would find  for each of  the two
$x_{0}$ values  alone. Further studies  are thus needed to  confirm to
what extent the  combination of only two scales  is universally robust
and effective enough, independently of the number of prominent spatial
scales  present in  the domain;  this  point will  be the  focus of  a
forthcoming work.

\section{Discussion}

The main result extracted from  the theoretical analysis reported here
is  that  MSRWs  with  only  two  movement  characteristic  scales  can
represent a mathematical  optimum (in terms of  MFPT minimization) for
random  search  strategies. This  has  been  proved by  checking  that
additional scales do  not allow to improve the optimum  achieved for 2
scales.  The optimal solution  outperforms both ballistic and L\'{e}vy
strategies  but  only for  specific  intervals  of the  characteristic
parameters $\tau_{i}$  and $w_{i}$ which depend  on the characteristic
scales of  the domain  (namely, $x_{0}$ and  $L$). In  particular, the
global  optimum turns  out  to  be given  by  $\tau_{1}  \gg L/v$  and
$\tau_{2} \approx  2 x_{0}/v$, which  can be intuitively  justified in
terms of optimizing the tradeoff between exploring for faraway targets
and  exploiting  nearby  resources.  While  the  theoretical  analysis
provided here  has been  restricted to  the one-dimensional  case (for
which an  exact solution for  the MFPT  is attainable), we  think that
these arguments are  generally valid and so we expect  them to hold in
higher dimensions, and probably in  more complicated situations as for
instance in biased searches \cite{metzler14} too.

In the  context of animal  foraging, the fact that  fine-tuned 2-scale
random-walks  outperform   L\'{e}vy  walks  represents   a  convenient
extension  of  the  L\'{e}vy   flight  paradigm  from  the  completely
uninformed scenario to  that in which domain scales  are (partially or
completely) available to  the organism. In the  uninformed case, where
the  characteristic  scales  of  the search  problem  are  unknown,  a
scale-free  strategy  represents  a  convenient  (albeit  sub-optimal)
solution. However,  in cognitive systems search  optimization programs
should be adjustable on the basis  of accumulated evidence. As we show
here, 2-scale  walks would be  optimal provided that the  searcher has
previous available information  (at least some crude  guess) about the
values of the scales $x_{0}$ and $L$. 
Let us stress that we are considering that such prior guess about target distances is limited (by the searcher cognitive capacity) or not informative enough (e.g. landscape noise, insufficient cumulative evidences) to set up a deterministic search strategy; so, the random-walk hypothesis is still meaningful. 
Accordingly, as more  and more information about the
domain scales becomes integrated by the searcher we should observe a tendency
towards a reduction (and an  adjustment) in the number (and magnitude)
of movement scales used, respectively. This process should go on up to
the point where  barely one or two scales  would persist. Furthermore,
we   note  that   for   the  extreme   case   of  perfectly   informed
(deterministic)  walkers no  characteristic search scales  at all  would be
necessary since  in that case  the search process is  plainly directed
towards the target.
%Note that two-scale exploratory behavior, for example, has been observed in the nematode {\it  C.elegans} \cite{srivastava09,salvador14}).   

Our  results  add then  some  new  dilemmas  and perspectives  on  the
fundamental problem  of what  biological scales  could be  relevant in
terms of a  program driving animal paths to  enhance foraging success.
Within  the  uninformed  scenario, Weierstrassian  Walks  involving  a
relatively low number  of scales in a geometric  progression have been
proposed  as   an  efficient  mechanism  to   implement  L\'{e}vy-like
trajectories  \cite{reynolds14}.   This  itself  builds  on  the  more
general  idea of  reproducing power-law  paths through  a hierarchical
family of  random walks  \cite{hughes81}.  These  Weierstrassian Walks
provide, due to its relative simplicity, a promising approach to bring
together the  ideas from the  L\'{e}vy flight paradigm and  those from
MSRWs \cite{benhamou14}, although we stress that many
alternative  Markovian  embeddings  for  power-laws do  exist  in  the
literature  \cite{kupferman04,goychuk09}.  Within  this  context,  the
existence of a  correlation between the number of  movement scales and
the informational  gain we  suggest here may  pose new  challenges for
experimentalists and  data miners. For  example, provided that  we can
conveniently interpret  animal trajectories in terms  of a combination
of scales, can we infer  something about the informational capacity of
the  individual  from the  number  of  scales  observed and  from  the
relation   between   their   values?    How   can   we   differentiate
informationally-driven scales  from the internally-driven  ones? While
we  are not  yet in  position to  provide a  definite answer  to these
questions,  we  expect  that  the  ideas provided  in  this  work  can
stimulate  research  in  this  line and  can  assist  experimentalists
towards new  experimental designs  for a  better understanding  of the
interplay between  animal foraging, landscape scales,  and information
processing.

\textbf{Acknowledgements.} This research  has been partially supported
by Grants No. FIS 2012-32334 (VM, DC), SGR 2013-00923 (VM, FB, DC) and the the Human Frontier Science Program RGY0084/2011 (FB). EPR acknowledges CNPq and FACEPE.

\newpage

\end{document}